\documentclass[aps,prc,showpacs,twocolumn,superscriptaddress]{revtex4-1}
\usepackage{amsmath,amsfonts,amssymb,amsthm,float,mathastext, array, booktabs, tabularx, mathtools}
\usepackage{booktabs}
\usepackage{graphicx}
\graphicspath{{fig/}}
\usepackage{siunitx}
\usepackage{hyperref}
\hypersetup{hidelinks,colorlinks=false,breaklinks=true,urlcolor=ocre}
\usepackage{dcolumn}
\usepackage{bm}
\usepackage[english]{babel}
\usepackage[autostyle]{csquotes}
\usepackage{color}
\begin{document}
\preprint{APS/123-QED}

\title{The Decay Q Value of Neutrinoless Double Beta Decay Revisited}

\author{D.-M. Mei} 
 \email{Corresponding author.\\Email: Dongming.Mei@usd.edu}
\affiliation{Department of Physics, The University of South Dakota, Vermillion, SD 57069, USA}
\author{W.-Z. Wei}
\affiliation{Department of Physics, The University of South Dakota, Vermillion, SD 57069, USA}

\date{\today}

\begin{abstract}
An earlier publication "The implication of the atomic effects in neutrinoless double beta (0$\nu\beta\beta$) decay" written by Mei and Wei has motivated us to compare the decay Q value ($Q_{\beta\beta}$) derived from the decay of the parent nucleus to the daughter nucleus with the two ejected beta particles in the final state to the $Q_{\beta\beta}$ directly derived from the decay of the initial neutral atom to the final state of double-ionized daughter ion with the two ejected beta particles in the final state. We show that the results are the same, which is the mass-energy difference ($\Delta Mc^2$) subtracted by the total difference of the atomic electron binding energy ($\Delta E_{b}$) between the ground states of initial and final neutral atoms. We demonstrate that $\Delta Mc^2$ is the sum of $Q_{\beta\beta}$ and the atomic relaxation energy ($\Delta E_{b}$) of the atomic structure after the decay. Depending on the atomic relaxation time, the release of the atomic binding energy may not come together with the energy deposition of the two ejected beta particles. 

\end{abstract}

\maketitle

Neutrinos take a key role in understanding the universe~\cite{aat}. The discovery of neutrino oscillation indicates that neutrinos have mass~\cite{superk}. This has motivated physicists to postulate new properties of neutrinos~\cite{moh}, which have created a possible connection between the observed asymmetry of matter over antimatter in our universe~\cite{apa}. If neutrinos are Majorana particles~\cite{majorana}, this means that neutrinos are their own anti-particles. This Majorana nature of neutrinos, if confirmed, might offer an explanation of the prevalence of matter over anti-matter~\cite{bar}. The only experimentally feasible way to answer whether neutrinos are Majorana particles is to search for neutrinoless double-beta (0$\nu\beta\beta$) decay~\cite{fra, agi}, a proposed form of rare nuclear decay. The experimental signature is the decay Q value from this decay process as explored by many experiments~\cite{gerda, mjd, exo, cuore}.  

In an earlier paper~\cite{meiwei}, Mei and Wei derived the decay Q value from the decay of the parent nucleus to the daughter nucleus with the two ejected beta particles in the final state. In this case, a 0$\nu\beta\beta$ decay process can be described as below~\cite{fra}:
\begin{equation}
\label{eq:react}
^{A}_{Z}X_{N}\longrightarrow^{A}_{Z+2}X_{N-2}^{'} + e^{-} + e^{-},
\end{equation}
where $^{A}_{Z}X_{N}$ is the mother nucleus in the initial state while $^{A}_{Z+2}X_{N-2}^{'}$ is the daughter nucleus in the final state. The two electrons in the final state are the two ejected beta particles as the result of this nuclear decay process. The decay Q value, which is defined as the available kinetic energy for the two ejected beta particles, for this nuclear decay process can be calculated using the mass difference in the rest frame between the initial and final products~\cite{meiwei}:
\begin{equation}
\label{eq:react1}
Q_{\beta\beta} = [m_{N}(^{A}_{Z}X) - m_{N-2}(^{A}_{Z+2}X^{'}) - 2m_{e}]c^{2},
\end{equation}
where $m_{N}(^{A}_{Z}X)$ stands for the mass of the mother nucleus in the rest frame while $m_{N-2}(^{A}_{Z+2}X^{'})$ represents the mass of the daughter nucleus in the rest frame. $m_{e}$ is the mass of electron in the rest frame. $c$ is the speed
of light. The mass of a nucleus is related to the corresponding mass of the neutral atom, which is denoted as $M(^{A}_{Z}X)$, through the following relation:
\begin{equation}
\label{eq:ee}
    M(^{A}_{Z}X)c^2 = m_{N}(^{A}X)c^2+Zm_ec^2 - \sum_{i = 1}^{Z}B_{i},
\end{equation}
where $B_{i}$ stands for the atomic binding energy of the $i$th electron. Note that the sign of the binding energies in Eq.~\ref{eq:ee} must be taken as positive values. Therefore, the atomic binding energies are all positive values in this work and the previous work by Mei and Wei~\cite{meiwei}. Replacing the masses of the  mother and daughter nuclei in Eq.~\ref{eq:react1} using the corresponding atomic masses in the neutral form expressed in Eq.~\ref{eq:ee}, by rearranging the terms, $Q_{\beta\beta}$ is found to be:
\begin{equation}
\label{eq:eq4}
Q_{\beta\beta} = [M(^{A}_{Z}X) - M(^{A}_{Z+2}X^{'})]c^{2} -  [\sum_{i=1}^{Z+2}B_{i} -  \sum_{i=1}^{Z}B_{i}].
\end{equation}
As can be seen in Eq.~\ref{eq:eq4}, $Q_{\beta\beta}$ is given by two terms. The first term,
\begin{equation}
\Delta Mc^2 = [M(^{A}_{Z}X)-M(^{A}_{Z+2}X^{'})]c^2,
\end{equation}
is the mass-energy difference between the atomic masses of the mother and daughter atoms in the neutral form, and the second term,
\begin{equation}
\Delta E_{b}= [\sum_{i=1}^{Z+2}B_{i} -  \sum_{i=1}^{Z}B_{i}]
\end{equation}
represents the difference of the total atomic binding energy of the mother and daughter atoms in the neutral form. Therefore, $Q_{\beta\beta}$ can be written as:
\begin{equation}
    \label{eq:add1}
    Q_{\beta\beta} = \Delta Mc^2 - \Delta E_{b}.
\end{equation}
It is important to point out that this is a standard definition for the decay Q value, which is similar to a normal beta decay process where the decay Q value is expressed as~\cite{gas, kra}:
\begin{equation}
 \label{eq:eq5}
   Q_{\beta} = [M(^{A}_{Z}X) - M(^{A}_{Z+1}X^{'})]c^{2} -  [\sum_{i=1}^{Z+1}B_{i} -  \sum_{i=1}^{Z}B_{i}], 
\end{equation}

Considering that 0$\nu\beta\beta$ decay in a nucleus occurs very fast in a level of picoseconds, the atomic structure may have no time to respond to the change of the nuclear charge right after the decay. The daughter nucleus would be surrounded by the atomic electrons from the mother atom. In this case, the daughter atom is a double-ionized ion. The decay Q value can be directly derived from the decay of the ground state of the mother atom to the double-ionized daughter ion with the two ejected beta particles in the final state. This process can be expressed as:
\begin{equation}
\label{eq:atom}
 ^{A}_{Z}X\longrightarrow^{A}_{Z}X^{''++} + e^{-} + e^{-},   
\end{equation}
where $^{A}_{Z}X$ is the parent atom in the neutral form and $^{A}_{Z}X^{"++}$ is the double-ionized daughter ion in which the daughter nucleus is surrounded by the parent atomic electrons right after the decay. Therefore,  the atomic member, $Z$, is the same as the parent atom when the nuclear charge is altered by two units in the final state of the nucleus. The decay Q value for this decay is calculated as:
\begin{equation}
 \label{eq:atom1}  
 Q_{\beta\beta} = M(^{A}_{Z}X)c^2 - M(^{A}_{Z}X^{''++})c^2-2m_{e}c^2,
\end{equation}
where $M(^{A}_{Z}X)$ is the mass of the parent atom in the neutral form and $M(^{A}_{Z}X^{''++})$ is the mass of the double-ionized daughter ion. Note that $M(^{A}_{Z}X^{''++})$ is related to the mass of the daughter nucleus through the following relation:
\begin{equation}
\label{eq:atom2}
M(^{A}_{Z}X^{''++})c^2 = m_{N-2}(^{A}_{Z+2}X^{'})c^2 + Zm_{e}c^2 - \sum_{i=1}^{Z}B_{i}, 
\end{equation}
where $m_{N-2}(^{A}_{Z+2}X^{'})c^2$ is the mass of the daughter nucleus, $Zm_{e}$ represents the total mass of the orbital electrons from the parent atom and $\sum_{i=1}^{Z}B_{i}$ stands for the total atomic binding energy from the parent atom. Since the mass of the daughter nucleus can be calculated using the mass of the daughter atom in the neutral form using Eq.~\ref{eq:ee}, thus,  $m_{N-2}(^{A}_{Z+2}X^{'})c^2$ can be expressed as:
\begin{equation}
    \label{eq:atom3}
    m_{N-2}(^{A}_{Z+2}X^{'})c^2 = M(^{A}_{Z+2}X^{'})c^2 - (Z+2)m_{e}c^2 + \sum_{i=1}^{Z+2}B_{i}, 
\end{equation}
where $M(^{A}_{Z+2}X^{'})c^2$ is the mass of the daughter atom in the neutral form, $(Z+2)m_{e}c^2$ is the total mass of the orbital electrons surrounding the daughter nucleus, and 
$\sum_{i=1}^{Z+2}B_{i}$ represents the total atomic binding energy of the daughter atom in the neutral form. Putting Eq.~\ref{eq:atom3} into Eq.~\ref{eq:atom2} and rearranging the terms, $M(^{A}_{Z}X^{''++})c^2$ can be expressed as:
\begin{equation}
    \label{eq:atom4}
    M(^{A}_{Z}X^{''++})c^2 = M(^{A}_{Z+2}X^{'})c^2 - 2m_{e}c^2 + \sum_{i=1}^{Z+2}B_{i} - \sum_{i=1}^{Z}B_{i}.
\end{equation}
Replacing $M(^{A}_{Z}X^{''++})c^2$ in Eq.~\ref{eq:atom1} using Eq.~\ref{eq:atom4}, 
 one obtains the following:
\begin{equation}
    \label{eq:atom5}
    Q_{\beta\beta} = M(^{A}_{Z}X)c^2 - M(^{A}_{Z+2}X^{'})c^2 - [\sum_{i=1}^{Z+2}B_{i} - \sum_{i=1}^{Z}B_{i}].
\end{equation}
Eq.~\ref{eq:atom5} is the decay Q value directly derived from the decay of the ground state of the mother atom in the neutral form to the double-ionized daughter ion as described in Eq.~\ref{eq:atom}. Note that Eq.~\ref{eq:atom5} is exactly the same as Eq.~\ref{eq:eq4}, which is the decay Q value derived from the decay of the parent nucleus to the daughter nucleus with the two ejected beta particles in the final state as described in Eq.~\ref{eq:react}. Since the decay Q value is defined as the mass difference between the initial state and the final state products, the decay Q value is expected to be the same regardless of how it is derived. 

From Eq.~\ref{eq:add1}, it is apparent that the decay Q value for 0$\nu\beta\beta$ is equal to the mass difference of the neutral atoms in the initial and final states subtracted by the difference of the total atomic electron binding energy between the mother and daughter atoms. This indicates that the calculation of $Q_{\beta\beta}$ must take into account the difference of the total atomic binding energy between the mother and daughter neutral atoms. As long as $Q_{\beta\beta}$ is calculated by using the mass-energy difference between the ground state of mother and daughter neutral atoms, it is required by the energy conservation as shown in Eq.~\ref{eq:ee} that $Q_{\beta\beta}$ = $\Delta Mc^2$ - $\Delta E_{b}$. 

\begin{figure}[htp!!!]
\includegraphics[width=7cm]{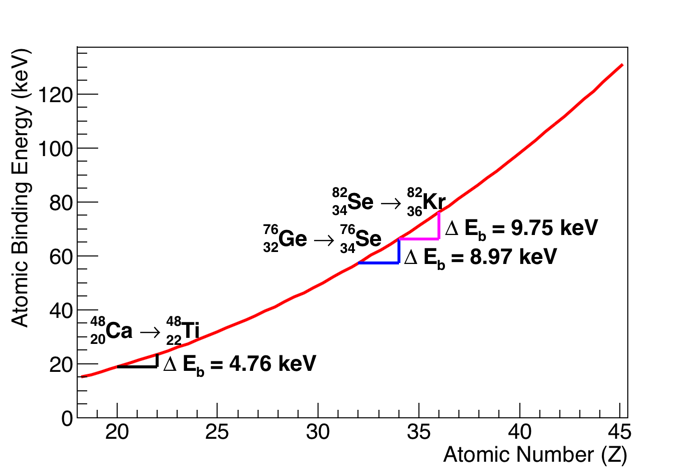}
\caption[]{
The electron binding energy as a function of atomic number $Z$ and the labelled electron binding energy difference for $^{48}_{20}Ca \rightarrow ^{48}_{22}Ti$, $^{76}_{32}Ge \rightarrow ^{76}_{34}Se$, and $^{82}_{34}Se \rightarrow ^{82}_{36}Kr$.
}
\label{fig:fig1}
\end{figure}

\begin{figure}[htp!!!]
\includegraphics[width=7cm]{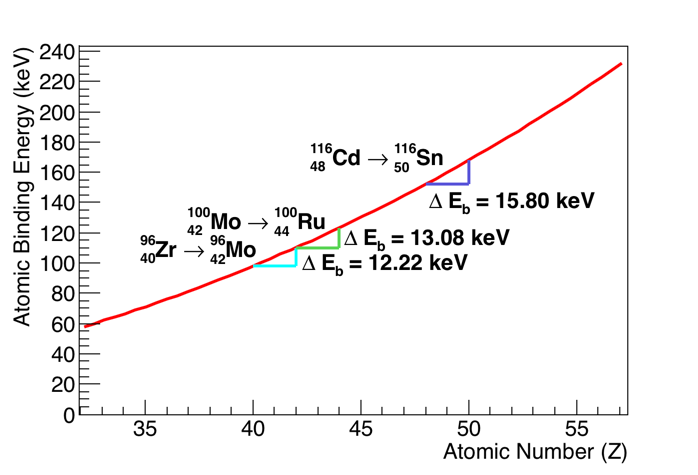}
\caption[]{
The electron binding energy as a function of atomic number $Z$ and the labelled electron binding energy difference for $^{96}_{40}Zr \rightarrow ^{96}_{42}Mo$, $^{100}_{42}Mo \rightarrow ^{100}_{44}Ru$, and $^{116}_{48}Cd \rightarrow ^{116}_{50}Sn$.
}
\label{fig:fig2}
\end{figure}

\begin{figure}[htp!!!]
\includegraphics[width=7cm]{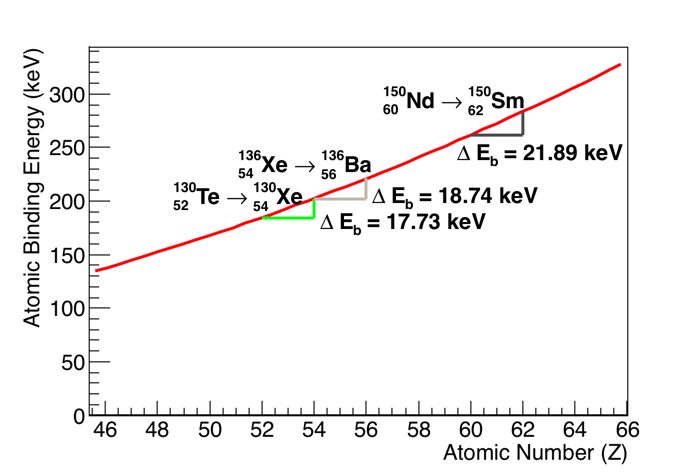}
\caption[]{
The electron binding energy as a function of atomic number $Z$ and the labelled electron binding energy difference for $^{130}_{52}Te \rightarrow ^{130}_{54}Xe$, $^{136}_{54}Xe \rightarrow ^{136}_{56}Ba$, and $^{150}_{60}Nd \rightarrow ^{150}_{62}Sm$.
}
\label{fig:fig3}
\end{figure}

To calculate $\Delta E_{b}$, one must know the total electron binding energy for a given atomic system. Since this quantity cannot be easily measured, a good approximation is given by Lunney, Perrson, and Thibault~\cite{lpt}:
\begin{equation}
    \label{eq:lpt}
    B_{e}(Z) = 14.4381Z^{2.39} + 1.55468\times10^{-6}Z^{5.35} eV,
\end{equation}
where $B_{e}(Z)$ represents the total electron binding energy for a given atomic system. Therefore, $\Delta E_{b}$ can be expressed as: 
\begin{equation}
    \label{eq:lpt1}
    \Delta E_{b} = B_{e} (Z+2) - B_{e} (Z),
\end{equation}
for any given 0$\nu\beta\beta$ decay process. 
Figures~\ref{fig:fig1},~\ref{fig:fig2}, and ~\ref{fig:fig3} display the calculated electron binding energy as a function of $Z$ according to Eq.~\ref{eq:lpt} for nine 0$\nu\beta\beta$ decay candidates and their corresponding decay daughters. The difference in the total electron binding energy, $\Delta E_{b}$, is labelled respectively for the nine 0$\nu\beta\beta$ decay processes. 
 As can be seen in Figures~\ref{fig:fig1},~\ref{fig:fig2}, and ~\ref{fig:fig3}, the difference in the total electron binding energy, $\Delta E_{b}$, ranges from a minimum of 4.76 keV for $^{48}_{20}Ca \rightarrow ^{48}_{22}Ti$ to a maximum of 21.89 keV for $^{150}_{60}Nd \rightarrow ^{150}_{62}Sm$. This will result in a significant contribution to the calculation of $Q_{\beta\beta}$ for a given 0$\nu\beta\beta$ decay process using Eq.~\ref{eq:add1}. Therefore, $\Delta E_{b}$ cannot be ignored. 
 
The decay Q value used by the current experiments~\cite{gerda, mjd, exo, cuore} is just taken as the mass-energy difference of the initial and the final state atoms in the neutral form. This refers to a decay of the following form:
\begin{equation}
    \label{eq:eq14}
    ^{A}_{Z}X\longrightarrow^{A}_{Z+2}X^{'},
\end{equation}
where $^{A}_{Z}X$ and $^{A}_{Z+2}X^{'}$ are the neutral atoms from the initial and the final states. This does not represent 0$\nu\beta\beta$ decay, which is expected to have two beta particles present in the final state. Therefore, the decay Q value derived from this decay form, $Q_{\beta\beta}$ = $M(^{A}_{Z}X)c^2$ - $M(^{A}_{Z+2}X^{'})c^2$, is not the decay Q value for 0$\nu\beta\beta$ decay. 
From Eq.~\ref{eq:add1}, one obtains the following:
\begin{equation}
    \label{eq:add2}
    \Delta Mc^2 = Q_{\beta\beta} + \Delta E_{b}.
\end{equation}
$\Delta Mc^2$ is the sum of the decay Q value for 0$\nu\beta\beta$ decay ($Q_{\beta\beta}$) and the relaxation energy ($\Delta E_{b}$) of the initial atomic structure after the decay. It is critical to distinguish the decay Q value, which is the available energy for the two ejected beta particles in the final state for 0$\nu\beta\beta$ decay, and the relaxation energy of the initial atomic structure. The former is the experimental signature for 0$\nu\beta\beta$ decay while the latter is the release of the atomic binding energy difference after the decay. The release of the atomic binding energy difference depends on the relaxation time of the atomic structure after the decay. This relaxation time is not well understood for 0$\nu\beta\beta$ decay since it has not been observed yet. The atomic relaxation time in Rydberg blockaded-$\Lambda$ atoms through spontaneous decay~\cite{qiao} may shed light on the atomic relaxation time of 0$\nu\beta\beta$ decay. In the recent paper written by Qiao et al.~\cite{qiao}, the atomic relaxation time is calculated using the atomic number and the decay rate as below:
\begin{equation}
    \label{eq:eq15}
    t_{r} = \frac{2Z}{\gamma_{rd}}\sqrt[2Z]{2\pi Z}(1-\frac{2}{\sqrt{Z}}),
\end{equation}
where $t_{r}$ is the atomic relaxation time, $\gamma_{rd}$ is the decay rate. 
If one uses $Z$ (the atomic number) and $\gamma_{rd}$ (the decay rate) for 0$\nu\beta\beta$ decay, the relaxation time of the atomic orbits can be very long. In particular, the lifetime of the double-ionized daughter ion in the final state can be very long. This is because the thermalized two ejected beta particles cannot be easily captured by the double-ionized daughter ion due to the repulsive Coulomb forces from large numbers of electrons surrounding the daughter nucleus. These repulsive forces provide a stabilizing energy balance between the two thermalized electrons and the positively charged daughter nucleus. An indicator of long-lived ions from $\beta$ decay is a good example of the above theory. The $^{42}$K ions, which are from the $\beta$ decay of $^{42}$Ar, produced uniformly throughout the Ar volume, are transported near the detectors by convective flow in the G{\sc{erda}} detector~\cite{gerda}, where ion collection onto the detector surfaces is aided by the stray electric fields from the biased detector surface and any unshielded high voltage components. The observed enhancement of $\gamma$ rays from the $\beta$ decay of $^{42}$K ion in the G{\sc{erda}} experiment indicates that the lifetime of $^{42}$K ions can be very long. Thus, it is reasonable to assume that the release of the atomic binding energy after the 0$\nu\beta\beta$ decay may not come within the detection time window for the energy deposition from the two ejected beta particles. 

Even if the release of the atomic binding after the 0$\nu\beta\beta$ decay is within the detection time window of the two ejected beta particles, the full relaxation of the atomic orbital structure is a multistep process, which results in a cascade of atomic radiation. Depending on the atomic number, the electron orbital configuration involved, and electron shells, the energy of emitted X-rays and/or Auger electrons is typically in the range from a few eV to a couple of keV. Such low energy X-rays or Auger electrons have a short range (nm to $\mu$m) to lose all energies. Therefore, the linear energy transfer (LET) is very high. A high LET means that the majority of energy loss is converted into the production of phonons. Depending on the detection technology, there is a chance that the majority of the atomic energy released by atomic radiation may not be detected.    

 In any case, for calculating $Q_{\beta\beta}$ for 0$\nu\beta\beta$ decay,  the difference in the total atomic binding energy between the mother and daughter neutral atoms must be taken into account because it is in the level of a few keV to more than 10 keV as also calculated in our earlier publication~\cite{meiwei}. This difference in the total atomic binding energy is larger than the detection energy window, which is so called the energy region of interest (ROI), used for several 0$\nu\beta\beta$ decay experiments such as G{\sc{erda}}~\cite{gerda}, M{\sc{ajorana}} D{\sc{emonstrator}}~\cite{mjd}, LEGEND~\cite{legend}, CUORE~\cite{cuore} and CUPID~\cite{cupid}. A narrow energy window can be used by these experiments because they possess excellent energy resolution. A narrow energy window allows these experiments to keep background events especially the 2$\nu\beta\beta$ decay events out of the ROI and thus have the potential to discover the 0$\nu\beta\beta$ decay process. However, based on the discussions in this work and the previous publication~\cite{meiwei}, the size of a narrow energy window, which can be used to search for 0$\nu\beta\beta$ decay, should be evaluated using the atomic binding energy difference between the mother and daughter atoms in the neutral form. Otherwise, 0$\nu\beta\beta$ decay experiments could miss the decay signature because the real decay Q value is out of the ROI, which is determined using $\Delta Mc^2$ used by the current experiments. This indicates that the study of the cascade of the atomic binding energy and its relaxation time becomes important for 0$\nu\beta\beta$ decay. 

The currently used decay Q value is the atomic mass difference of the neutral atoms from the initial and the final states. The PENNING-trap technique has been used to measure directly the atomic mass differences between the neutral atoms~\cite{eron} in the initial and the final states. These measured values are 4267.98 keV for $^{48}$Ca $\rightarrow^{48}$Ti~\cite{kwia}, 2039.04 keV for $^{76}$Ge $\rightarrow^{76}$Se~\cite{raha}, 2997.73 keV for $^{82}$Se $\rightarrow^{82}$Kr~\cite{linc}, 3355.85 keV for $^{96}$Zr $\rightarrow^{96}$Mo~\cite{guly}, 3034.40 keV for $^{100}$Mo $\rightarrow^{100}$Ru~\cite{raha}, 2813.50 keV for $^{116}$Cd $\rightarrow^{116}$Sn~\cite{raha1}, 2527.01 keV for $^{130}$Te $\rightarrow^{130}$Xe~\cite{raha1}, 2457.83 keV for $^{136}$Xe $\rightarrow^{136}$Ba~\cite{reds}, and 3367.72 keV for $^{150}$Nd $\rightarrow^{150}$Sm~\cite{kolh}, respectively. These values should be corrected for the total difference in the atomic binding energy between the initial and the final neutral atoms as shown in Figures~\ref{fig:fig1}, \ref{fig:fig2}, and \ref{fig:fig3}.

Note that the decay Q-value can be measured using 2$\nu\beta\beta$ decay spectra as conducted by the NEMO collaboration for $^{100}$Mo in R. Arnold et al.~\cite{arno}. However, the spectra measured using the calorimeter technique should correct for the energy loss of the emitted electrons in order to determine the end-point energy for the 2$\nu\beta\beta$ decay process. Thus, this end-point is the decay Q-value for 0$\nu\beta\beta$.

In conclusion, we derive the decay Q value for 0$\nu\beta\beta$ decay in the forms of the parent nucleus decay into the daughter nucleus with the two ejected beta particles in the final state and the parent neutral atom decay into the double-ionized daughter ion with the two ejected beta particles in the final state, respectively. We show that the decay Q value is the same in both cases. If one uses the atomic masses of the mother and daughter atoms in the neutral form to calculate the decay Q value, $Q_{\beta\beta}$ = $\Delta Mc^2$ - $\Delta E_{b}$, which is different from the current value ($\Delta Mc^2$) used in the experiments. For example, this difference ($\Delta E_{b}$) can be as large as 8.97 keV for 0$\nu\beta\beta$ decay in $^{76}$Ge, 17.73 keV for 0$\nu\beta\beta$ in $^{130}$Te, and 18.74 keV for 0$\nu\beta\beta$ in $^{136}$Xe, respectively, as shown in Figures~\ref{fig:fig1}, \ref{fig:fig2}, and \ref{fig:fig3}. We emphasize that $\Delta Mc^2$ is not the decay Q value for 0$\nu\beta\beta$ decay and is the sum of the decay Q value and the relaxation energy of the total atomic binding energy difference between the mother and daughter atoms in the neutral form. This atomic relaxation after the decay may not come within the detection time window depending on the atomic relaxation time, which is not well understood for 0$\nu\beta\beta$ decay since it has not been observed yet. Therefore, it is worth pointing out that the current and planned 0$\nu\beta\beta$ decay experiments may need to re-evaluate the ROI, which should be larger than the current ROI, as we suggested in the previous publication~\cite{meiwei}. 

This work was supported in part by NSF OISE 1743790, DE-SC0004768, and a governor's research center supported by the State of South Dakota. The authors would like to thank Dr. Christina Keller for a careful reading of the manuscript. 
%
%


\end{document}